\documentclass[aps,pra,reprint,superscriptaddress]{revtex4-2}
\usepackage{graphicx}
\usepackage{braket}
\usepackage{amsmath}
\usepackage{amsthm}
\usepackage{amssymb}
\usepackage{dsfont}
\usepackage[colorlinks,
            linkcolor=blue,
            anchorcolor=black,
            citecolor=blue,
			urlcolor=blue]{hyperref}
			
\usepackage[ruled]{algorithm2e}
\usepackage{makecell}
\usepackage{array}
\usepackage{color}

\newtheorem{definition}{Definition}

\begin{document}


\title{Determining the upper bound of code distance of quantum stabilizer codes through Monte Carlo method based on fully decoupled belief propagation}


\author{Zhipeng Liang}
\author{Zicheng Wang} 
\author{Zhengzhong Yi}
\author{Yulin Wu}
\author{Chen Qiu}
\author{Xuan Wang}
\email[]{zhengzhongyi@cs.hitsz.edu.cn}
\email[]{wangxuan@cs.hitsz.edu.cn}
\affiliation{Harbin Institute of Technology, Shenzhen. Shenzhen, 518055, China}


\date{\today}

\begin{abstract}
	Code distance is an important parameter for quantum stabilizer codes (QSCs). Directly precisely computing it is an NP-complete problem. However, the upper bound of code distance can be computed by some efficient methods. In this paper, employing the idea of Monte Carlo method, we propose the algorithm of determining the upper bound of code distance of QSCs based on fully decoupled belief propagation. Our algorithm shows high precision — the upper bound of code distance determined by the algorithm of a variety of QSCs whose code distance is known is consistent with actual code distance. Besides, we explore the upper bound of logical $X$ operators of Z-type Tanner-graph-recursive-expansion (Z-TGRE) code and Chamon code, which is a kind of XYZ product code constructed by three repetition codes. The former is consistent with the theoretical analysis, and the latter implies the code distance of XYZ product codes can very likely achieve $O(N^{2/3})$, which supports the conjecture of Leverrier \textit{et al.}.

\end{abstract}


\maketitle

\section{Introduction}
\label{1}

Quantum computing can solve certain problems that classical computing cannot solve within limited resources and time\cite{shor1994algorithms}. However, as the basic unit of quantum computing, qubit is susceptible to environmental noise, which makes final computation results go wrong, making it difficult to fulfil the potential of quantum computing. Fortunately, in 1995, Shor\cite{PhysRevA.52.R2493} and Steane\cite{PhysRevLett.77.793} proposed that we can use quantum error correcting codes to solve this problem, which makes reliable quantum computing become possible.

QSCs\cite{gottesman1997stabilizer} are an important class of quantum error correcting codes. When designing a new QSC, it’s necessary to precisely determine its code distance, since the code distance determines the number of physical qubits that can be reliably corrected. Generally, there are three methods to precisely determine the code distance of a QSC — theoretical proof method, linear programming method and exhaustive search. For some QSCs, such as planar surface code\cite{bravyi1998quantum,PhysRevA.86.032324}, XZZX surface code\cite{bonilla2021xzzx} and $D$-dimensional ($D\ge2$) toric codes\cite{kitaev2003fault,hamma2005string,castelnovo2008topological,breuckmann2016local}, their code distance can be determined by the properties of their topological structure. For QSCs with systematic construction method, such as hypergraph product code\cite{tillich2013quantum}, we can also theoretically prove their code distance by using the construction method. For some QSCs whose code distance cannot be theoretically proved by the above two method, one can compute their code distance by linear programming method\cite{landahl2011fault} and exhaustive search, but the time complexity of these two methods is exponential\cite{landahl2011fault}. Although it has been proven theoretically that exactly or approximately computing the code distance of QSCs is an NP-complete problem\cite{kapshikar2023hardness}, the upper bound of code distance can be determined by more efficient methods, such as Monte Carlo method.

Monte Carlo method\cite{rubinstein2016simulation} is widely employed in many fields to efficiently and approximately solve complex problems through stochastic simulation. In quantum error correcting field, the method can be used to rapidly determine the upper bound of code distance of some QSCs\cite{Panteleev2021degeneratequantum,bravyi2023high}. The basic procedure of using Monte Carlo method to determine the code distance of a given QSC is as follows. Performing multiple error correcting simulation trials on the given code, then recording the weight of the error correcting results which are logical operators, and finally finding the minimum value among them, which is the upper bound of code distance of the given QSC. We emphasize that the method has no strict restriction on the type of decoding algorithm, as long as the used decoding algorithm is applicable to the tested QSC and has good performance on it, the method is capable of determining the upper bound of code distance. However, when designing a new QSC, we usually don’t know which decoding algorithm is applicable to it. Therefore, a highly general, high-performance and fast decoding algorithm is very important for using Monte Carlo method to determine the upper bound of code distance.

Belief Propagation (BP) is a low time complexity decoding algorithm, and can be applied to all QSCs. However, its disadvantage is that its error correcting performance on QSCs is usually quite poor due to short cycles\cite{mackay2004sparse}. In Ref. \cite{Panteleev2021degeneratequantum}, researchers shows that the performance of BP can be improved by ordered statistics decoding (OSD)\cite{fossorier1995soft} when BP fails to converge. Besides, researches show that binary BP combined with OSD (BP-OSD) is applicable to many Calderbank-Shor-Steane (CSS)\cite{calderbank1998quantum} quantum low-density parity check (QLDPC) codes \cite{roffe2020decoding}, and has achieved pretty good error correcting performance on these codes. Therefore, BP-OSD is a highly general, high-performance and fast decoding algorithm. However, for non-CSS codes, traditional binary BP cannot provide a satisfactory error-correcting performance especially under $Y$-biased noise even combined with OSD.

The decoding algorithm we use here is fully decoupled BP combined with OSD (FDBP-OSD) proposed in our previous work\cite{yi2023improved}. There are two major advantages of this decoding algorithm, 1) it has satisfactory decoding accuracy not only for CSS codes, but also for non-CSS codes and 2) compared with the traditional binary BP\cite{babar2015fifteen,poulin2008iterative}, FDBP has higher convergence rate and decoding accuracy. 

In this paper, employing the idea of Monte Carlo method, we propose the algorithm  of determining the upper bound of code distance of QSCs based on FDBP-OSD. First, we use this algorithm to determine the upper bound of the code distance of a variety of QSCs whose code distance is known, such as planar surface code, XZZX surface cod and toric code to verity the effectiveness of the algorithm. Our algorithm shows high precision — their upper bound of code distance determined by the algorithm is consistent with actual code distance. Second, we explore the Z-TGRE codes proposed in our previous work\cite{yi2022quantum}, and the minimum weight of logical $X$ operators determined by the algorithm is consistent with our theoretical analysis. Third, we explore the upper bound of code distance of Chanmon code, which is the XYZ product\cite{leverrier2022quantum} of three repetition codes with block lengths $n_1$, $n_2$, and $n_3$ and whose code distance has not been understood well so far. Our results show that when $n_1=n_2=n_3=L$, its upper bound of code distance is $2L$, and when $n_1=L-1$, $n_2=L$, $n_3=L+1$, its upper bound of code distance is $L(L-1)$. This results implies that the code distance of XYZ product codes can very likely achieve $O(N^{2/3})$\cite{leverrier2022quantum}. We emphasize that, since FDBP-OSD is a highly general, high-performance and fast decoding algorithm for both CSS and non-CSS codes, when designing new QSCs and their code distance is hard to compute, the algorithm is a useful method to quickly determine the upper bound of their code distance.

The rest of paper is organized as follows. In Sect. \ref{2}, we introduce some preliminaries, including quantum stabilizer code, Z-TGRE code and XYZ product code. Sect. \ref{3} introduces how to determining the upper bound of code distance of QSCs through Monte Carlo method based on FDBP-OSD. The simulation results are presented in Sect. \ref{4}. In Sect. \ref{5}, we conclude our work.

\section {Preliminaries}
\label{2}

\subsection{Quantum stabilizer code}
\label{2.1}
In this section, we briefly introduce the basic concept of QSCs. QSCs are an important class of quantum error correcting codes, which are the analogue of classical linear codes in quantum information field.

The code space $\mathcal{Q}_C$ of an $[[n,k,d]]$ QSC $C$ is a $2^k$-dimensional subspace of the Hilbert space $\mathcal{H}_2^{\otimes n}$, which is stabilized by an Abelian stabilizer group $\mathcal{S}\in\mathcal{G}_n$, where $\mathcal{G}_n$ is the $n$-fold tensor product of single qubit Pauli group $\mathcal{G}_1=\left\{\pm I,\ \pm iI,\ \pm X,\ \pm iX,\ \pm Y,\ \pm iY,\ \pm Z,\ \pm iZ\right\}$. More precisely, 
\begin{equation}
	Q_C=\{\ket{\varphi}\in (\mathcal{H}^2 )^{\otimes n}:S\ket{\varphi}=\ket{\varphi},\forall S\in \mathcal{S}\}
\end{equation}
The stabilizer group $\mathcal{S}$ can be generated by $k$ independent Pauli operators on $n$ qubits $S_1,\cdots,S_{n-k}\in\mathcal{G}_n$, namely, $\mathcal{S}=\langle S_1,\cdots,S_{n-k} \rangle$. Giving a set of stabilizer generators $S_1,\cdots,S_{n-k}$ of code $C$ is equivalent to explicitly giving the code space $\mathcal{Q}_C$.

The error syndrome $s=\left(s_1,\cdots,s_{n-k}\right)$ of an error $E\in\mathcal{G}_n$ is a binary vector, where $s_i=1$ if $E$ anti-commutes with $S_i$ and $0$ otherwise.

In quantum information theory, researchers usually use two bits to represent Pauli $I$, $X$, $Y$, $Z$ operators, namely,

\begin{equation}
	I\rightarrow(0,0),\ X\rightarrow(1,0),\ Z\rightarrow (0,1),\ Y\rightarrow (1,1)
\end{equation}

In this way, any operator $E\in\mathcal{G}_n$ acting on $n$ qubits can be represented as a binary vector $\textbf{e}=\left(\textbf{e}_x\mid\textbf{e}_z\right)$. Based on this Pauli-to-GF$\left(2\right)$ isomorphism, the binary parity-check matrix $H$ of a $[[n,k,d]]$ QSC is a block matrix with dimension $\left(n-k\right)\times2n$, which consists of two $\left(n-k\right)\times n$ binary matrices $H_x$ and $H_z$, namely,
\begin{equation}
	H=(H_x\mid H_z)
\end{equation}
And the syndrome $s$ of $E$ is computed by
\begin{equation}
	\label{syndrome}
	s=(H_x\cdot \textbf{e}_z + H_z\cdot \textbf{e}_x)\ mod\ 2
\end{equation}

For a QSC, if its stabilizer generators can be divided into two parts, each of which only contains either $X$-type or $Z$-type Pauli operators, it is a CSS code, otherwise it is a non-CSS code. In this way, the parity check matrix of a CSS code can be written as
\begin{equation}
	H = \begin{pmatrix}
		H_x& \textbf{0}\\
		\textbf{0} & H_z
	\end{pmatrix}
\end{equation}
where $H_x$ and $H_z$ both have $n$ columns and the commutation condition $H_xH_z^T=\textbf{0}$ satisfies.

The weight of an operator $P\in\mathcal{G}_n$ is defined as the number of qubits on which it acts nontrivially, and we use notation $wt\left(\cdot\right)$ to denote it. For instance, $wt\left(I_1X_2Y_3Z_4\right)=3$.

The logical operators of a QSC C are the set of operators in $\mathcal{G}_n$ which commute with all elements in $\mathcal{S}$ but are not in $\mathcal{S}$. More precisely, the logical operators are the elements of $\mathcal{C}(\mathcal{S})/\mathcal{S}$, where $\mathcal{C}(\mathcal{S})$ is the centralizer of $\mathcal{S}$ defined as $\mathcal{C}(\mathcal{S})={P\in\mathcal{G}_n:SP=PS,\ \forall S\in\mathcal{S}}$. For a $[[n,k,d]]$ QSC, we can find k pairs of logical operators ${({\bar{X}}_j,{\bar{Z}}_j)}_{j=1,\cdots,k}$ such that ${\bar{X}}_i{\bar{Z}}_j={(-1)}^{\delta_{ij}}{\bar{Z}}_j{\bar{X}}_i$, where $\delta$ is the Kronecker delta, which means for the same pair of logical operators ${\bar{X}}_j,{\bar{Z}}_j$, they are anti-commute, but they commute with other pairs of logical operators. We can see that $\mathcal{C}(\mathcal{S})=S_1,\cdots,S_{n-k},X_1,Z_1,\cdots,X_k,Z_k$. The code distance $d$ is defined as the minimum weight of logical operators, namely,
\begin{equation}
	d=\min\limits_{L\in\mathcal{C}(\mathcal{S})/\mathcal{S}}{wt(L)}
\end{equation}

The code distance $d$ of a QSC $C$ determines the number of qubits, $t$, up to which code $C$ can reliably correct, namely $t=\left\lfloor\frac{d-1}{2}\right\rfloor$. In general, there are three methods to precisely compute the code distance of a QSC — theoretical proof method, linear programming method\cite{landahl2011fault} and exhaustive search. However, only a part of QSCs’ code distance can be theoretically proved. For some QSCs whose code distance cannot be theoretically proved, one can compute their code distance by linear programming method or exhaustive search. However, the time complexity of these two methods is exponential. Moreover, it has been proven that exactly or approximately computing the code distance of a QSCs is an NP-complete problem\cite{kapshikar2023hardness}.

\subsection{Z-type Tanner-graph-recursive-expansion code}
\label{2.2}
\begin{figure*}[htbp]
	\centering
	\includegraphics[width=0.75\textwidth]{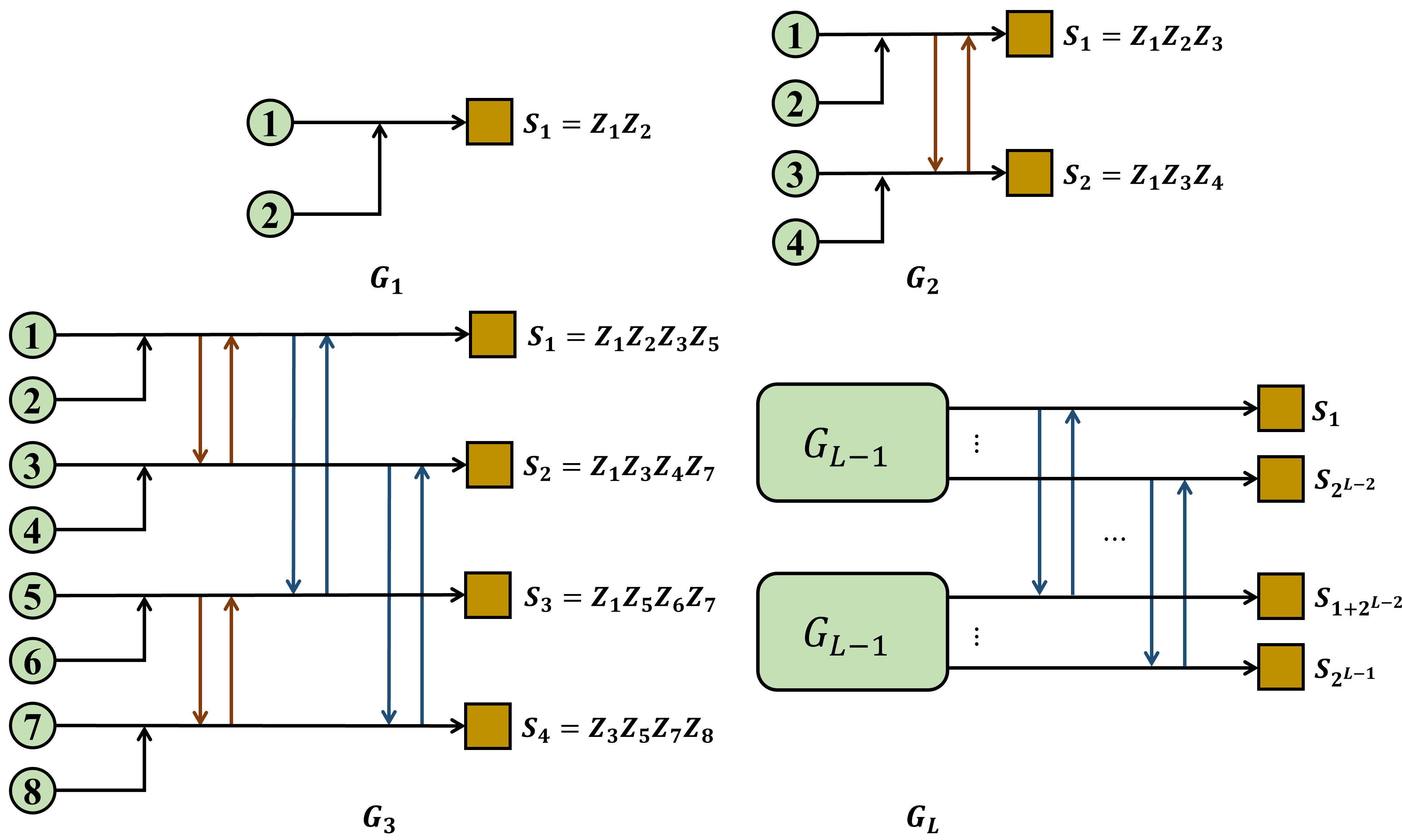}
	\caption{The Tanner graph recursive expansion of Z-TGRE codes. The arrow means the corresponding variable node (the qubit) it starts from will join in the corresponding the check node (the stabilizer) it ends with. The variable is numbered from 1 to $N=2^L$. $G_1$ is the primal Tanner graph used to recursive expansion. $G_2$ is the expanded Tanner graph by the recursive expansion of two primal Tanner graphs $G_1$. $G_3$ is the expanded Tanner graph by the recursive expansion of two $G_2$. $G_L$ is the expanded Tanner graph by recursive expansion of two $G_{L-1}$.}
	\label{Z-TGRE}
\end{figure*}

In our previous work \cite{yi2022quantum}, we propose a new class of quantum stabilizer codes named Z-TGRE code, which is obtained by recursively expanding Tanner graph, and have constant coding rate of 0.5, but can only correct Pauli $X$ and $Y$ errors. The way to expand the Tanner graph of Z-TGRE code is shown in Fig. \ref{Z-TGRE}.

Our theoretical analysis shows that for a Z-TGRE code with code length $N=2^L$, if $L$ is an even, the minimum weight of logical operators $\bar{X}$ is $\log N$. If $L$ is an odd, the minimum weight of logical operators $\bar{X}$ is $\log N+1$. Readers can see \cite{yi2022quantum} for more detail.

\subsection{XYZ product code}
\label{2.3}

XYZ product \cite{leverrier2022quantum} is a three-fold variant of the hypergraph product code construction, which yields non-CSS QSCs. To better understand XYZ product code, we should first introduce how to describe a CSS code in terms of chain complexes and hypergraph product code.

A chain complex $\mathfrak{C}$ of length $L$ is a collection of $L+1$ vector spaces $C_0 \cdots C_L$ and $L$ linear maps $\partial_i:\ C_i \longrightarrow C_{i+1}$ ($0\le i\le L-1$), namely,
\begin{equation}
	\mathfrak{C}=\left(C_0\stackrel{\partial_0}{\longrightarrow}C_1\stackrel{\partial_1}{\longrightarrow}\cdots\stackrel{\partial_{i-1}}{\longrightarrow}C_i\stackrel{\partial_{i}}{\longrightarrow}C_{i+1}\stackrel{\partial_{i+1}}{\longrightarrow}\cdots\stackrel{\partial_{L-1}}{\longrightarrow}C_L\right)
\end{equation}
which satisfies $\partial_{i+1}\partial_i=0$.

A chain complex $\mathfrak{C}$ with length 2 naturally define a CSS code $\mathcal{C}(\mathfrak{C})$, namely
\begin{equation}
	\mathfrak{C}=\left(\mathbb{F}_{2}^{m_z}\stackrel{H_z^T}{\longrightarrow}\mathbb{F}_{2}^{n}\stackrel{H_x}{\longrightarrow}\mathbb{F}_{2}^{m_x}\right)
\end{equation}
where the commutation condition $H_xH_z^T=\textbf{0}$ is naturally satisfied. 

It is easy to see that a classical code $\mathcal{C}=ker H$ similarly corresponds to a chain complex with length 1, namely, 
\begin{equation}
	\mathbb{F}_{2}^{n}\stackrel{H}{\longrightarrow}\mathbb{F}_{2}^{m}
\end{equation}

Hypergraph product is using two classical codes $\mathcal{C}_1=ker H_1$ and $\mathcal{C}_2=ker H_2$ to construct a CSS code $\mathcal{C}$ (where $H_i$, $i\in\{1,2\}$, are the parity check matrices of size $m_i\times n_i$ of codes $\mathcal{C}_i$), which corresponds to the following length-2 chain complex,
\begin{equation}
	\mathbb{F}_{2}^{m_1\times n_2}\xrightarrow{H_z^T}
	\mathbb{F}_{2}^{n_1\times n_2}\oplus\mathbb{F}_{2}^{m_1\times m_2}\xrightarrow{H_x}\mathbb{F}_{2}^{n_1\times m_2}
\end{equation}
where $H_x = \mathds{1}_{n_1}\otimes H_{2}\oplus H_{1}^{T}\otimes \mathds{1}_{m_2}$ and $H_z = (H_1^T\otimes \mathds{1}_{n_2}\oplus \mathds{1}_{m_1}\otimes H_{2})^T$.

\begin{figure*}[htbp]
	\centering
	\includegraphics[width=0.75\textwidth]{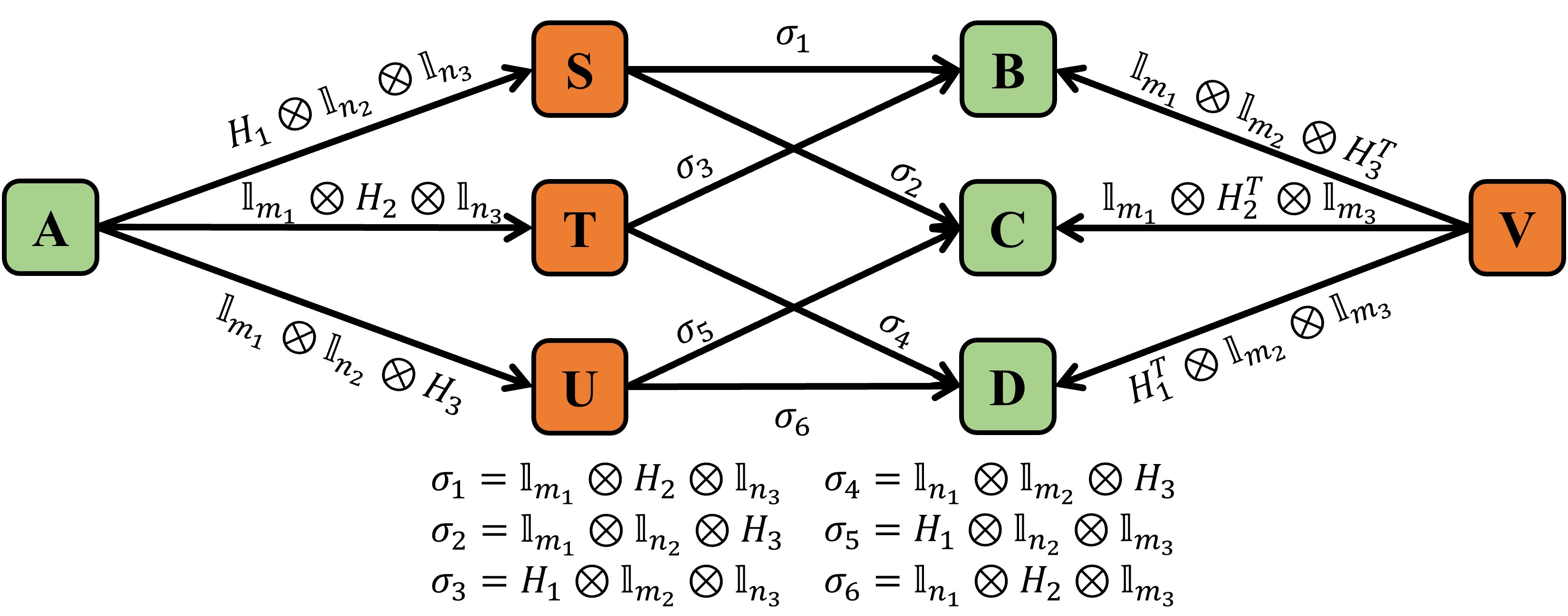}
	\caption{“Chain complex” representation of the XYZ product code. }
	\label{XYZ chain complex}
\end{figure*}

The XYZ product code construction is a variant of the hypergraph product code construction, which use three classical codes to construct a non-CSS code. Specifically, given three parity check matrices $H_i$ of size $m_i\times n_i$ $\left(i=1,2,3\right)$, the stabilizer generator matrix $\mathcal{S}$ of the corresponding XYZ product code is
\begin{widetext}
\begin{equation}
	\mathcal{S} = \begin{bmatrix}
		X^{\left(H_1\otimes\mathds{1}_{n_2}\otimes\mathds{1}_{n_3}\right)},Y^{\left(\mathds{1}_{m_1}\otimes H_2^T\otimes\mathds{1}_{n_3}\right)},Z^{\left(\mathds{1}_{m_1}\otimes\mathds{1}_{n_2}\otimes H_3^T\right)},I^{\left(m_1n_2n_3\times n_1m_2m_3\right)}\\
		Y^{\left(\mathds{1}_{n_1}\otimes H_2\otimes\mathds{1}_{n_3}\right)},X^{\left(H_1^T\otimes \mathds{1}_{m_2}\otimes\mathds{1}_{n_3}\right)},I^{\left(n_1m_2n_3\times m_1n_2m_3\right)},Z^{\left(\mathds{1}_{n_1}\otimes\mathds{1}_{m_2}\otimes H_3^T\right)}\\
		Z^{\left(\mathds{1}_{n_1}\otimes\mathds{1}_{n_2}\otimes H_3\right)},I^{\left(n_1n_2m_3\times m_1m_2n_3\right)},X^{\left(H_1^T\otimes \mathds{1}_{n_2}\otimes\mathds{1}_{m_3}\right)},Y^{\left(\mathds{1}_{n_1}\otimes H_2^T\otimes \mathds{1}_{m_3}\right)}\\
		I^{\left(m_1m_2m_3\times n_1n_2n_3\right)},Z^{\left(\mathds{1}_{m_1}\otimes\mathds{1}_{m_2}\otimes H_3\right)},Y^{\left(\mathds{1}_{m_1}\otimes H_2\otimes\mathds{1}_{m_3}\right)},X^{\left(H_1\otimes\mathds{1}_{m_2}\otimes\mathds{1}_{m_3}\right)}
	\end{bmatrix}
\end{equation}
\end{widetext}
Each row of $\mathcal{S}$ corresponds to a stabilizer generator. Here the notation $\mathcal{P} = P^{H}\ (P\in\{X,Y,Z\})$ denotes a Pauli tensor, which means for any entry of matrix $H$, if it is a $1$, $\mathcal{P}$ has a Pauli operator $P$ at the corresponding position, and a identity operator $I$ otherwise. Fig. \ref{XYZ chain complex} gives the "chain complex" representation of the corresponding XYZ product code.
$\textbf{A}$, $\textbf{B}$, $\textbf{C}$ and $\textbf{D}$ are vector spaces which index the qubits and $\textbf{S}$, $\textbf{T}$, $\textbf{U}$ and $\textbf{V}$ are vector spaces which index the stabilizer generators, namely, 
\begin{equation}
	\begin{aligned}
		&\textbf{A}\in\mathbb{F}_{2}^{n_1\times n_2\times n_3},\ \textbf{B}\in\mathbb{F}_{2}^{m_1\times m_2\times n_3},\\
		&\textbf{C}\in\mathbb{F}_{2}^{m_1\times n_2\times m_3},\ \textbf{D}\in\mathbb{F}_{2}^{n_1\times m_2\times m_3}
	\end{aligned}	
\end{equation}
and
\begin{equation}
	\begin{aligned}
	&\textbf{S}\in\mathbb{F}_{2}^{m_1\times n_2\times n_3},\ \textbf{T}\in\mathbb{F}_{2}^{n_1\times m_2\times n_3},\\
	&\textbf{U}\in\mathbb{F}_{2}^{n_1\times n_2\times m_3},\ \textbf{V}\in\mathbb{F}_{2}^{m_1\times m_2\times m_3}
	\end{aligned}
\end{equation}
It can be seen that the code length is $N=n_1 n_2 n_3+m_1 m_2 n_3+m_1 n_2 m_3+n_1 m_2 m_3$. As for the dimension of a XYZ product code, we need to find the number of independent stabilizer generators, and readers can see Ref. \cite{leverrier2022quantum} for more detail.

In Ref. \cite{leverrier2022quantum}, Leverrier \textit{et al.} conjecture that this code family includes codes whose minimum distance is $O(N^{2/3})$. However, no one has proved it. The simplest instance of XYZ product code is Chamon code \cite{chamon2005quantum}, which is the XYZ product of 3 repetition codes with block lengths $n_1$, $n_2$, and $n_3$, whose code distance has not been understood well. In Sect. \ref{4}, employing Monte Carlo method based on FDBP-OSD, we show that, when $n_1=n_2=n_3=L$, its upper bound of code distance is $2L$, and when $n_1=L-1$, $n_2=L$, $n_3=L+1$, its upper bound of code distance is $L(L-1)$. The results implies that the code distance of XYZ product code can very likely achieve $O(N^{2/3})$.

\section {Determining the upper bound of code distance}
\label{3}
Although computing the code distance of QSCs is a NP-complete problem, the upper bound of code distance can be determined by some efficient methods, such as Monte Carlo method.
This section first introduces the general idea of determining the upper bound of code distance through Monte Carlo method, then briefly reviews FDBP decoding algorithm proposed in Ref. \cite{yi2023improved}, which has higher convergence rate and decoding accuracy compared with traditional binary BP. Finally, this section introduces how to employ FDBP combined with OSD (FDBP-OSD) to determine the upper bound of code distance based on Monte Carlo method.

\subsection {Monte Carlo method}
\label{3.1}
\begin{figure}[htbp]
	\centering
	\includegraphics[width=0.45\textwidth]{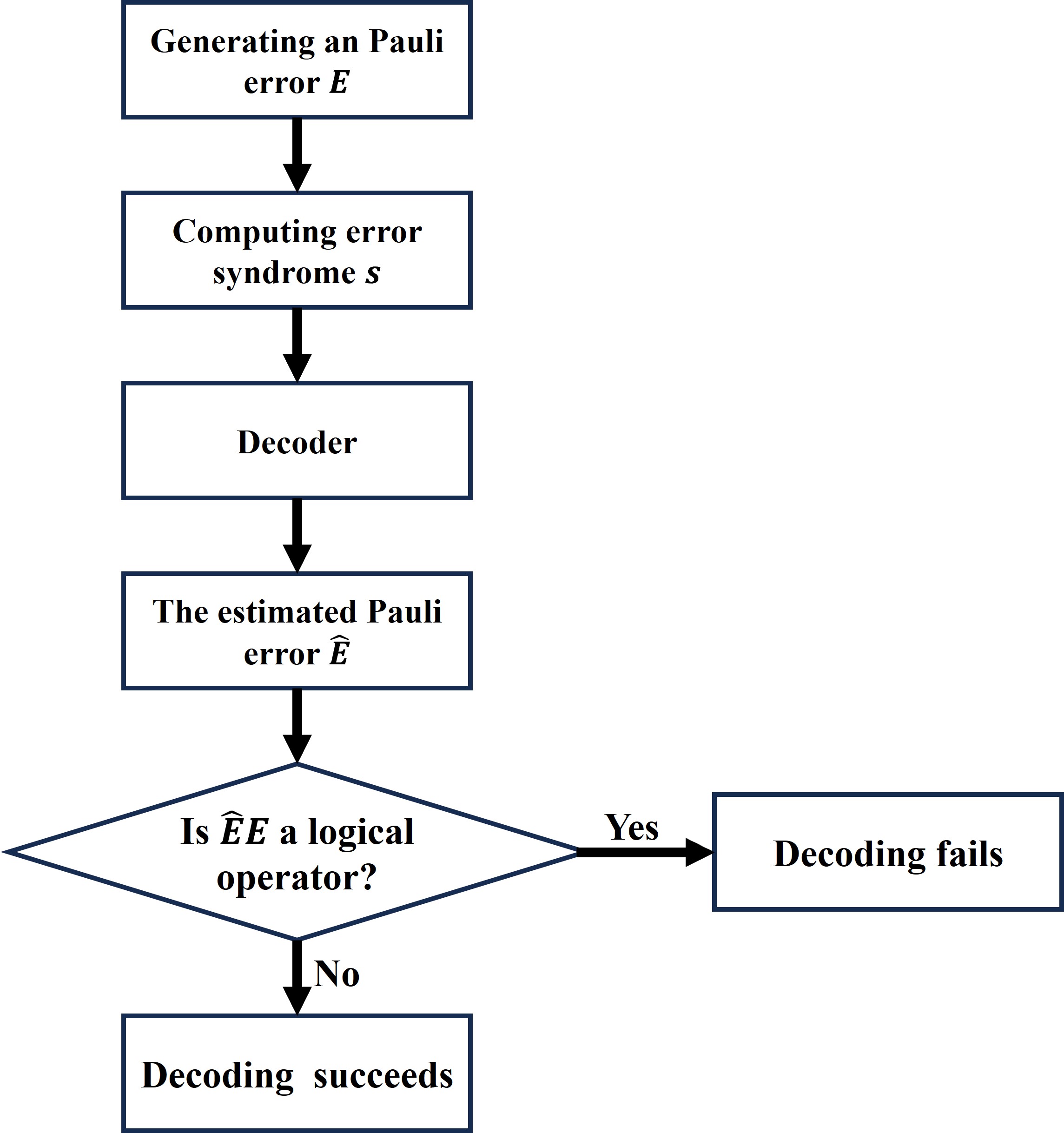}
	\caption{The procedure of quantum error correction simulation.}
	\label{FlowGraph}
\end{figure}

To understand the general idea of determining the upper bound of code distance through Monte Carlo method, it’s necessary to comprehend the procedure of quantum error correction simulation shown in Fig. \ref{FlowGraph}. As shown in Fig. \ref{FlowGraph}, the first step is to randomly generate a Pauli error $E$ and compute the corresponding error syndrome $s$. Then $s$ is input into a decoder, and subsequently the decoder output an estimated Pauli error $\hat{E}$ whose corresponding syndrome is also $s$. The last step is to compute $E\hat{E}$, if it is a logical operator, decoding is failed. Otherwise it is a stabilizer, the decoding succeeds.

A good decoder should follow the idea of maximum likelihood decoding to make the probability of the estimated Pauli error $\hat{E}$ it output as high as possible, which means the weight of $\hat{E}$ should be as small as possible. In addition, in the low physical qubit error rate regime, the weight of $E$ should also be small. Thus, if $E\hat{E}$ is a logical operator, the probability that it is in the minimum-weight form should be higher than it isn’t, which is the key idea of determining the upper bound of code distance through Monte Carlo method.

One can see that this method has no strict restriction on the type of decoder, as long as it as applicable to the tested QSC. However, in some cases, especially when we design a new QSC, which decoder is applicable to it is not known. Thus, a decoder with high generality and performance is very important for this method. FDBP \cite{yi2023improved} combined with OSD is a highly general and high-performance decoder, thus we can employ it to determine the upper bound of code distance of QSCs.

\subsection {Fully decoupled belief propagation}
\label{3.2}
FDBP proposed in our previous work\cite{yi2023improved} is a kind of improved binary BP for QSCs, which has higher convergence rate and decoding accuracy compared with traditional binary BP \cite{babar2015fifteen,poulin2008iterative}. 

For traditional binary BP for QSCs, the symplectic representation of Pauli operators results that Pauli Y errors introduce the correlations between vectors $\textbf{e}_x$ and $\textbf{e}_z$, which decrease its performance. To deal with the correlations, a binary BP using the $X/Z$ correlations is proposed\cite{6874997}. However, this method is only applicable to CSS codes. For non-CSS codes, especially in $Y$-biased noise, even combined with OSD, the performance of traditional binary BP in symplectic representation is unsatisfactory.

To handle this problem, we propose FDBP decoding algorithm for QSCs, which can eliminate correlations between vectors $\textbf{e}_x$ and $\textbf{e}_z$ in symplectic representation introduced by Pauli $Y$ errors. Thus, it is applicable to both CSS and non-CSS codes. The FDBP decoding algorithm is based on the following three modifications. First, we represent single-qubit Pauli operators by three bits as shown in Definition \ref{Decoupling representation of Pauli operators}, which is called decoupling representation of Pauli operators, such that the correlations introduced by Pauli $Y$ errors are eliminated. Second, based on Definition \ref{Decoupling representation of Pauli operators}, the decoupled parity-check matrix of a QSC is obtained as shown in Definition \ref{Decoupled parity-check matrix}. Third, noticed that, in the decoupling representation $\textbf{e}=\left(\textbf{e}_x\prime\mid\textbf{e}_z\prime\mid\textbf{e}_y\prime\right)$ of a Pauli operator $E$ acting on $n$ qubits, the total number of $1$s of the $i$th $(i\le n)$, $\left(i+n\right)$th and $\left(i+2n\right)$th elements of e is no more than one, namely, $e_i+e_{i+n}+e_{i+2n}\le1$. Thus, this restraint condition should be taken into account, according to which message update and hard decision rules are modified, and finally FDBP is obtained. Our simulation results in Ref. \cite{yi2023improved} show that the convergence rate and decoding accuracy of FDBP are higher than those of traditional binary BP. Readers can see Ref. \cite{yi2023improved} for more details.
\begin{definition}[\textbf{Decoupling representation of Pauli operators}]\cite{yi2023improved}
	\label{Decoupling representation of Pauli operators}
	The representation which represents Pauli operators by the following mapping is called decoupling representation.
	\begin{equation}
		\begin{aligned}
			&I\rightarrow(0,0,0),\ X\rightarrow(1,0,0),\\
			&Z\rightarrow(0,1,0),\ Y\rightarrow(0,0,1)
		\end{aligned}	
	\end{equation}
	For a Pauli error $E$ acting on $n$ qubits, by the above mapping, its decoupling representation is a binary vector $\textbf{e}$ with size of $3n$, namely,
	\begin{equation}
		\textbf{e}=(\textbf{e}_x^{\prime}\mid\textbf{e}_z^{\prime}\mid\textbf{e}_y^{\prime})
	\end{equation}
	where $\textbf{e}_x^{\prime}$, $\textbf{e}_z^{\prime}$ and $\textbf{e}_y^{\prime}$ are all binary vectors with size of $n$. Taking $X_1Y_2Z_3$ as an example, the corresponding decoupling representation is $\textbf{e}=(\textbf{e}_x^{\prime}\mid\textbf{e}_z^{\prime}\mid\textbf{e}_y^{\prime})=(1\ 0\ 0\mid0\ 0\ 1\mid0\ 1\ 0 )$.
\end{definition}

\begin{definition}[\textbf{Decoupled parity-check matrix}]\cite{yi2023improved}
	\label{Decoupled parity-check matrix}
	Given an $[[n,k]]$ QLDPC code $C$ and the symplectic representation of its stabilizer generators $H=(H_x\mid H_z)$, the decoupled parity-check matrix of $C$ is
	\begin{equation}
		H_d=(H_z\mid H_x \mid(H_x\oplus H_z))
	\end{equation}
	whose dimension is $(n-k)\times 3n$ and where $\oplus$ denotes addition modulo 2. We can see that given the decoupling representation $\textbf{e}=(\textbf{e}_x^{\prime}\mid\textbf{e}_z^{\prime}\mid\textbf{e}_y^{\prime})$ of a Pauli error E and the decoupled parity-check matrix $H_d=(H_z\mid H_x \mid(H_x\oplus H_z))$ of a QSC, the error syndrome $s$ is
	\begin{equation}
		\textbf{s}=(H_d\cdot \textbf{e})\ mod\ 2
	\end{equation}
\end{definition}

\subsection {Algorithm}
\label{3.3}
Algorithm \ref{algorithm 1} shows how to employ FDBP-OSD to determine the upper bound of code distance. The key idea of the algorithm is to run FDBP-OSD many times in different physical qubit error rate such that the found upper bound of code distance will be closer to the real code distance. The noise model we use in simulation is depolarizing noise model, namely, given depolarizing error rate p, each of the Pauli operators $X$, $Y$ and $Z$ is applied to the qubit with probability $\frac{p}{3}$.
%
%
\begin{algorithm}[htbp]
	\caption{Determining the upper bound of code distance based on FDBP-OSD}
	\label{algorithm 1}
	\LinesNumbered
	\KwIn{decoupled parity-check matrix $H_d=(H_z\mid H_x \mid(H_x\oplus H_z))$ and logical operators $L=({\bar{X}}_1,{\bar{Z}}_1,\cdots,{\bar{X}}_k,{\bar{Z}}_k)$ of the tested $[[N,k]]$ QSCs,\\
		a sequence of physical qubit error rate $p_1,p_2,\cdots,p_m$,\\
		the number of trials $T$,\\
		code length $N$.}
	\KwOut{the upper bound of code distance $d_{up}$}
	$MinimumWeight = N$
	
	\For{$p\gets p_1\ to\ p_m$}{
	\For{$i\gets1\ to\ T$}{
	According to $p$ randomly generating an error vector $\textbf{e}=\left(\textbf{e}_x^{\prime}\mid\textbf{e}_z^{\prime}\mid\textbf{e}_y^{\prime}\right)$ which corresponds to a Pauli error.
	
	Computing the corresponding error syndrome $\textbf{s}=(H_d\cdot \textbf{e})\ mod\ 2$
	
	Decoding $\hat{\textbf{e}}=\textbf{FDBP-OSD}\left(\textbf{s}\right)$
	
	\If{$\hat{\textbf{e}}+\textbf{e}$ is the error vector of a logical operator}{
	\If{$weight(\hat{\textbf{e}}+\textbf{e}) < MinimumWeight$}{
	$MinimumWeight = weight(\hat{\textbf{e}}+\textbf{e})$
	}
	}
	}
	}
	$d_{up}= MinimumWeight$
\end{algorithm}

\section{Simulation results}
\label{4}
In this section, employing Algorithm \ref{algorithm 1}, we first compute the upper bound of the code distance of planar surface code\cite{bravyi1998quantum,PhysRevA.86.032324}, XZZX surface code\cite{bonilla2021xzzx} and toric codes\cite{kitaev2003fault}, whose code distance is known, to verify the effectiveness of the algorithm. Second, we perform simulations on the Z-TGRE codes proposed in our previous work\cite{yi2022quantum}. Third, we explore the upper bound of code distance of Chamon codes, which is the XYZ product of three repetition codes with block lengths $n_1$, $n_2$ and $n_3$, and whose code distance is still unknown.

\begin{figure*}[htbp]
	\centering
	\includegraphics[width=1\textwidth]{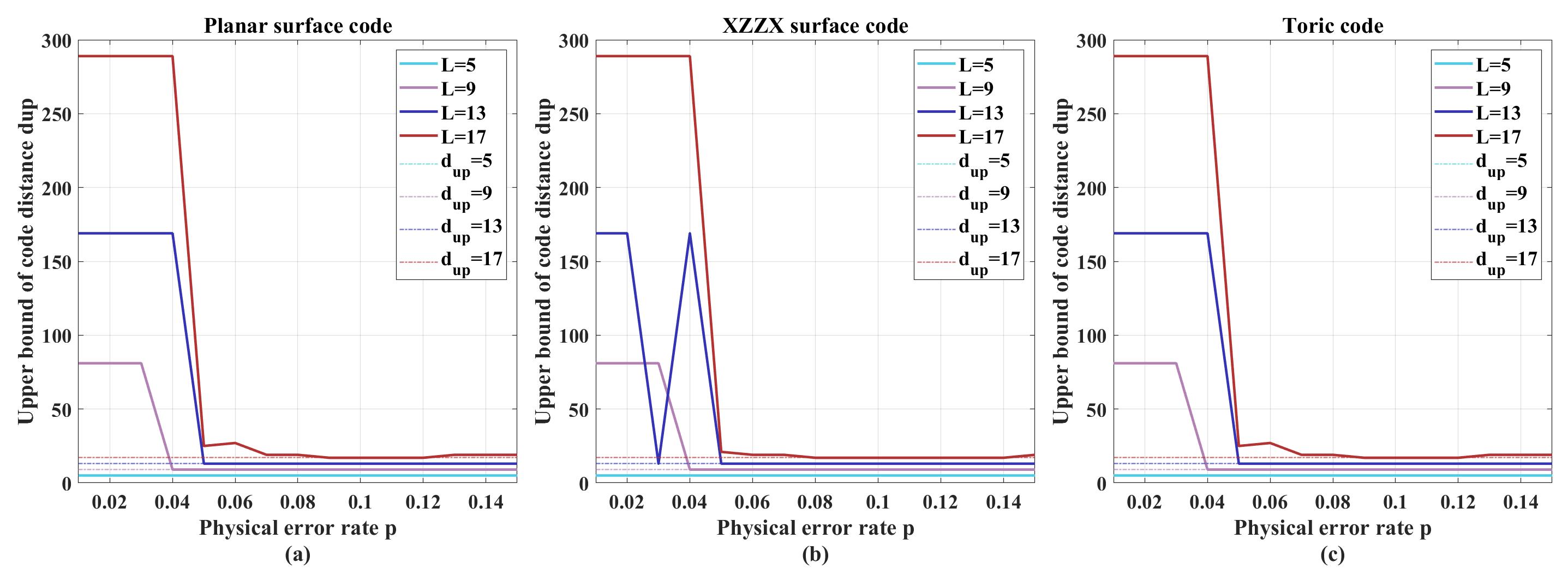}
	\caption{The upper bound of the code distance of (a) planar surface code, (b) XZZX surface code and (c) toric code respectively, which are determined by Algorithm \ref{algorithm 1} in different physical qubit error rate.}
	\label{SurfaceToricXZZX}
\end{figure*}

Fig. \ref{SurfaceToricXZZX} (a)$\sim$(c) shows the upper bound of the code distance of planar surface code, XZZX surface code and toric code respectively, which are determined by Algorithm \ref{algorithm 1} in different physical qubit error rate. It can be seen that the upper bound of the code distance of these codes found by Algorithm 1 is consistent with their code distance, which shows the high precision of this algorithm. It can be seen that in the low physical qubit error rate regime the upper bound of code distance found by Algorithm \ref{algorithm 1} is consistent with code length. The reason is that the upper bound of the code distance is initialized to code length in our simulations, and in the low physical qubit error rate regime all decoding trials  succeed due to the high performance of FDBP-OSD. Our simulation results show that in the error rate regime around $p=0.1$, it is much easier to obtain the upper bound of code distance that is closer to the real code distance.


In Ref. \cite{yi2022quantum}, we propose a new class of QSCs, Z-TGRE codes, with coding rate $\frac{1}{2}$, which can only correct pure Pauli X or Y errors. Table \ref{table_Z-TGRE_distance} shows the minimum weight of logical $X$ operators determined by Algorithm \ref{algorithm 1} (right column) of Z-TGRE codes with different code length, which is consistent with the theoretical analysis (middle column). 
\begin{table}[htbp]
	\begin{center}
		\caption{The minimum weight of logical $X$ operators $wt_{min}(\bar{X})$ (theoretical and found by Algorithm \ref{algorithm 1}) of Z-TGRE codes with code length $N$ from 4 to 512.}		
		\begin{tabular}{c|c|c}
			\hline
			Code length &$wt_{min}(\bar{X})$  &$wt_{min}(\bar{X})$\\
			$N$&theoretical& determined by Algorithm\ref{algorithm 1}\\
			\hline
			4 & 2 & 2 \\
			\hline
			8 & 4 & 4 \\
			\hline
			16 & 4 & 4 \\
			\hline
			32 & 6 & 6 \\
			\hline
			64 & 6 & 6 \\
			\hline
			128 & 8 & 8 \\
			\hline
			256 & 8 & 8 \\
			\hline
			512 & 10 & 10 \\
			\hline
		\end{tabular}
		\label{table_Z-TGRE_distance}
	\end{center}
\end{table}

Table \ref{Chamon_code_distance} shows that, for Chamon code, which is the XYZ product of 3 repetition codes with length $n_1$, $n_2$ and $n_3$, when $n_1=n_2=n_3=L$, its upper bound of code distance is $2L$, and when $n_1=L-1$, $n_2=L$ and $n_3=L+1$, its upper bound of code distance is $L(L-1)$. The results implies that the code distance of XYZ product codes can very likely achieve $O(N^{2/3})$, which supports the conjecture of Leverrier \textit{et al.}\cite{leverrier2022quantum}.

\begin{table}[htbp]
	\begin{center}
		\caption{The upper bound of code distance of Chamon codes, which is the XYZ product of 3 repetition codes with length $n_1$, $n_2$ and $n_3$.}		
		\begin{tabular}{c|c|c|c|c}
			\hline
			$n_1$ & \ $n_2$ & \ $n_3$ & code length $N$ & the upper bound of code distance \\
			\hline
			2 & 2 & 2 & 32 & 4 \\
			\hline
			3 & 3 & 3 & 108 & 6 \\
			\hline
			4 & 4 & 4 & 256 & 8 \\
			\hline
			5 & 5 & 5 & 500 & 10 \\
			\hline
			2 & 3 & 4 & 96 & 6 \\
			\hline
			3 & 4 & 5 & 240 & 12 \\
			\hline
			4 & 5 & 6 & 480 & 20 \\
			\hline
		\end{tabular}
		\label{Chamon_code_distance}
	\end{center}
\end{table}

\section {Conclusion}
\label{5}
Code distance is an important parameter of QSCs, since it determines the number of qubits that can be reliably corrected. However, precisely computing the code distance is an NP-complete problem, and sometimes we need to compute the upper bound of code distance instead, since it can be determined by more efficient methods. Monte Carlo method is to efficiently and approximately solve complex problems through stochastic simulation. In this paper, employing the idea of Monte Carlo method, we propose the algorithm of determining the upper bound of code distance of QSCs based on FDBP-OSD. First, The upper bound of code distance of planar surface code, XZZX surface code and toric code determined by the algorithm is consistent with their code distance, which shows effectiveness and high precision of the algorithm. Second, the minimum weight of logical $X$ operators determined by the algorithm of Z-TGRE codes is consistent with our theoretical analysis. Third, using this algorithm, we explore the code distance of Chamon code, which is the XYZ product of three repetition codes with length $n_1$, $n_2$ and $n_3$. Our results show that when $n_1=n_2=n_3=L$, its upper bound of code distance is $2L$, and when $n_1=L-1$, $n_2=L$ and $n_3=L+1$, its upper bound of code distance is $L(L-1)$. The results implies that the code distance of XYZ product codes can very likely achieve $O(N^{2/3})$, which supports the conjecture of Leverrier et al.\cite{leverrier2022quantum}. We emphasize that, since FDBP-OSD is a highly general, high-performance and fast decoding algorithm for both CSS and non-CSS codes, when designing new QSCs and their code distance is hard to compute, the algorithm is a useful method to quickly determine the upper bound of their code distance.

\section*{End Notes}
\subsection*{Acknowledgements}
 This work was supported by the Colleges and Universities Stable Support Project of Shenzhen, China (No.GXWD20220817164856008), the Colleges and Universities Stable Support Project of Shenzhen, China (No.GXWD20220811170225001) and Harbin Institute of Technology, Shenzhen - SpinQ quantum information Joint Research Center Project (No.HITSZ20230111).

\section*{Data Availability}
The data that support the findings of this study are available from the corresponding author upon reasonable request.

\bibliography{sn-bibliography}

\end{document}